\documentclass[twocolumn,showpacs,preprintnumbers,amsmath,amssymb]{revtex4}

\usepackage{epsfig}
\usepackage{dcolumn}
\usepackage{bm}

\begin{document}

\preprint{{\it Physical Review Letters}, In Press, arXiv:physics/0406063}

\title{Three species collisionless reconnection: Effect of O$^+$ on
magnetotail reconnection}

\author{M. A. Shay} \email{shay@glue.umd.edu}
\homepage{http://www.glue.umd.edu/~shay}
\author{M. Swisdak}
\affiliation{
Institute for Research in Electronics and Applied Physics\\
University of Maryland, College Park, MD, 20742
}

\date{\today}

\begin{abstract}
The nature of collisionless reconnection in a three-species plasma
composed of a heavy species, protons, and electrons is
examined. Besides the usual two length scales present in two-species
reconnection, there are two additional larger length scales in the
system: one associated with a ``heavy whistler'' which produces a
large scale quadrupolar out-of-plane magnetic field, and one
associated with the ``heavy Alfv\'en'' wave which can slow the
outflow speed and thus the reconnection rate. The consequences for
reconnection with O$^+$ present in the magnetotail are discussed. 
\end{abstract}

\pacs{Valid PACS appear here}

\maketitle

{\bf Introduction:} Recent studies of collisionless reconnection have
shown that the disparate masses of the ions and electrons lead to a
two-scale dissipation region near the x-line. The decoupling of the
ions from the magnetic field at larger scales than the electrons can
lead to whistler or kinetic Alfv\'en physics in the dissipation
region, whose quadratic dispersive characteristics can substantially
increase the reconnection
rate\cite{Biskamp95,Shay99,Birn01,Rogers01}. A
Sweet-Parker-like\cite{Vasyliunas75} analysis of this dissipation
region yields insight into the reconnection rate: $V_{in} \sim
(\delta/D) c_{At},$ where $V_{in}$ is the inflow speed, $\delta$ and
$D$ are the width and length of the dissipation region, and $c_{At}$
is the Alfv\'en speed just upstream from the dissipation region.

Many plasma systems have heavier species, in addition to protons and
electrons, which may play an important dynamical role: negatively
charged dust grains in astrophysical molecular clouds and the
interstellar medium, and O$^+$ in the Earth's
magnetosphere\cite{Shelley72}. Previous simulations of three-species
reconnection have focussed on O$^+$ in the
magnetosphere\cite{Winglee04,Birn04,Hesse04}, where the number density
of O$^+$ can sometimes exceed that of the
protons\cite{Peterson81,Kistler04a}. These simulations, however,
either did not find any effect of O$^+$ on reconnection due to the
small system size or did not spatially resolve the reconnection
boundary layers.

In this paper we present the first comprehensive study of basic
three-fluid reconnection showing through theory and simulation both
the effect of the heavy species on the reconnection rate and the
hierarchy of scales present in the microscale boundary layers. We find
that the usual two scales associated with collisionless two-fluid
reconnection ($d_i = c/\omega_{pi}, d_e = c/\omega_{pe}$) are instead
replaced by four scales. The inner two scales are associated with a
light whistler and a light Alfv\'en wave, which are very similar to
their two-fluid counterparts. At larger scales, however, a heavy
whistler and heavy Alfv\'en wave occur. The heavy whistler can occur
on scales much larger than a $d_i$ and thus gives rise to a much wider
quadrupolar out-of-plane magnetic field signature. Associated with
this magnetic field are parallel ion Hall currents, the analogue to
the light whistler electron currents. The higher O$^+$ mass substantially
slows the reconnection rate because the outflow speed from the x-line
is reduced from the usual proton Alfv\'en speed, $c_{Ai},$ to the much
slower heavy Alfv\'en speed, $c_{At}.$ 

{\bf Analytical Analysis:} We begin with the general three fluid
equations\cite{harold94} and first ignore electron inertia so that
${\bf E} = - {\bf V}_e/c \times {\bf B} - \nabla P_e/(n_e e).$ We
assume quasi-neutrality, $n_e = n_i + z_h n_h,$ where $n_i$ is the
light ion density, $n_h$ is the heavy species density, and $z_h$ is
the charge number of the heavy species. We ignore the displacement
current, ${\bf J} = (c/4\pi) \nabla \times {\bf B}.$ We normalize
length to $d_i = c\sqrt{m_i}/\sqrt{4\pi n_{i0} e^2}$ and time to
$\Omega_i^{-1} = m_i c/(e B_0),$ which gives the following equations:
\begin{equation}
\frac{\partial n_\alpha}{\partial t} = - \nabla \cdot \left( n_\alpha
{\bf V}_\alpha \right),\; \; \; \alpha = \{i,h\}
\label{dndt}
\end{equation}
\begin{equation}
\label{duidt}
n_i \frac{d {\bf V}_i}{dt} = 
z_h n_h ( {\bf V}_e - {\bf V}_h ) \times {\bf B} +
{\bf J} {\times} {\bf B} - 
\nabla{P_i} - \frac{n_i}{n_e} \nabla P_e
\end{equation}
\begin{equation}
\label{duhdt}
\hat{m}_h n_h \frac{d{\bf V}_h}{dt} = 
z_h n_h ({\bf V}_h - {\bf V}_e ) \times {\bf B} -
\nabla P_h -
\frac{z_h n_h}{n_e} \nabla P_e
\end{equation}
\begin{equation}
\label{dbdt}
\frac{\partial {\bf B}}{\partial t} = \nabla \times ( {\bf V}_e \times
{\bf B} ),
\end{equation}
where $n_e {\bf V}_e = n_i {\bf V}_i + z_h n_h {\bf V}_h - {\bf J},$
$d{\bf V}_\alpha/dt = (\partial/\partial t + {\bf V}_\alpha \cdot
\nabla ) {\bf V}_\alpha,$ $\hat{m}_h = m_h/m_i,$ ${\bf J} = \nabla
\times {\bf B},$ and $P_\alpha = T_\alpha n_\alpha$, where $T_\alpha$
is assumed to be an unchanging spatial constant (isothermal
approximation).

Although reconnection is a highly nonlinear process, much information
about its nature can be gleaned from a linear analysis, for it is bent
field line waves which ultimately accelerate the plasma away from the
x-line. We write each variable as ${\bf f}({\bf x},t) = {\bf f}_0({\bf
x}) + \tilde{\bf f} e^{i ({\bf k} \cdot {\bf x} - \omega t)}.$ Beginning
with a uniform ${\bf B}_0$ with no initial velocities, we proceed to
linearize Eqns.~\ref{dndt}-\ref{dbdt} and assume that ${\bf k}
\parallel {\bf B}_0$ for simplicity. The sound waves with ${\bf k}
\parallel$ (${\bf V}_i$ and ${\bf V}_h$) decouple from the magnetic
waves, leaving the following dispersion relation:
\begin{eqnarray}
\label{dispersion}
& \frac{\omega^3}{\Omega_i^3} \pm 
\frac{\omega^2}{\Omega_i^2}
   \left[ \frac{z_h n_h}{n_e} - \frac{\Omega_h}{\Omega_i} \frac{z_h n_h}{n_e} 
   + \frac{\Omega_h}{\Omega_i} - k^2 d_s^2 \right] \nonumber\\
& - \frac{\omega}{\Omega_i} k^2 d_s^2
  \left[ 1 + \frac{\Omega_h}{\Omega_i} \right] \mp
k^2 d_s^2 \frac{\Omega_h}{\Omega_i}  = 0,
\end{eqnarray}
where $\Omega_i = e B_0/(m_i c),$ $\Omega_h = e z_h B_0/(m_h c),$ $d_s
= d_i \sqrt{n_i/n_e} = c \sqrt{m_i}/\sqrt{4 \pi n_e e^2},$ $d_h =
c\sqrt{m_h}/\sqrt{4\pi n_h z_h^2 e^2},$ $n_e = n_i + z_h n_h,$ and all
densities, $n,$ are equilibrium quantities. This equation is fully
general and can apply to any plasma with electrons, ions, and a
third species.

Balancing the second and fourth terms and taking the limit $\omega \ll
\Omega_h \lesssim \Omega_i$ and $k^2 d_h^2 \ll 1$ yields the largest
scale or global Alfv\'en wave: $\omega = \pm k c_{At},$ where $c_{At}
= B/\sqrt{4 \pi (m_h n_h + m_i n_i)}.$ In order for the heavy species
to slow the global Alfv\'en wave appreciably, it is necessary for $m_h
n_h \gg m_i n_i.$

\begin{figure}
\epsfig{file=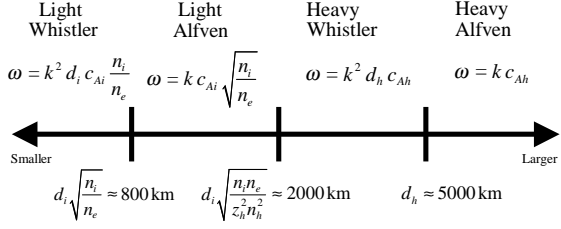,width=3.5in,clip=}
\caption{(below line) The non-ideal length scales present in
three-fluid reconnection and numbers for typical magnetotail lobes
with O$^+$ present\cite{Kistler04a} ($n_i = 0.05\, {\rm cm}^{-3},$
$n_h/n_i = 0.64$). (above line) Waves and dispersion relations at each
scale range.}
\label{bent}
\end{figure}

Taking the limit of $\Omega_i \gg \Omega_h$, $\omega \gg \Omega_h,$
and $m_h n_h \gg m_i n_i$ yields the high frequency dispersion:
\begin{equation}
\label{highfreq}
\frac{\omega^2}{\Omega_i^2} \pm
\frac{\omega}{\Omega_i}\left( \frac{z_h n_h}{n_e} - k^2 d_s^2 \right) -
k^2 d_s^2 = 0
\end{equation}
For $|z_h| n_h/n_e \leq 1,$ this equation produces the light whistler,
the light Alfv\'en and the heavy whistler waves shown in
Fig.~\ref{bent}, where $d_h^2 = c^2 m_h/(4 \pi n_h z_h^2 e^2),$ and
$c_{Ah} = B/\sqrt{4 \pi n_h m_h}.$ The existence of the heavy whistler
wave has been noted in electron-positron-dusty plasmas\cite{Shukla97}
and electron-proton-dusty plasmas\cite{Rudakov01,ganguli04}, but was
not applied to reconnection.
Taking $k^2 d_s^2 n_e/|z_h n_h| \gg 1,$ and then equating the first and
second terms yields the light whistler with $\omega = \pm k^2 d_i
c_{Ai} (n_i/n_e),$ where $c_{Ai} = B/\sqrt{4 \pi m_i n_i}$ is the proton
Alfv\'en speed. Equating the first and third term yields the light
Alfv\'en wave with $\omega = \pm k c_{Ai} \sqrt{n_i/n_e}.$ The
transition between these two waves occurs when $k^2 d_s^2 \sim 1.$
Both of these waves are very similar to their two-species analogues
except $n_i$ has been replaced with $n_e.$ 

Taking $k^2 d_s^2 n_e/|z_h n_h| \ll 1$ and equating the first and
third terms also yields the light Alfv\'en wave. Equating the second
and third term yields the heavy whistler wave with $\omega = \pm k^2
d_h c_{Ah}.$ The transition between these two waves occurs at $k^2
d_i^2 n_i n_e/(z_h^2 n_h^2) \sim 1.$ The heavy whistler requires
$\omega \gg \Omega_h$ so that the heavy species is unmoving, but the
ion inertia term in Eq.~\ref{duidt} is negligible. Thus, the wave is
characterized by frozen-in protons and electrons that flow together
and act as a massless fluid, but because $n_i \neq n_e,$ this net flow
is a current. This wave transitions to the heavy Alfv\'en wave at $k
d_h \sim 1$ with $\omega = k c_{Ah}$ in this limit.

The two scale structure of the dissipation region in a collisionless
two-fluid plasma\cite{Shay98a} ($d_e$ and $d_i$) has now been replaced
with four scales: the three scales in Fig.~\ref{bent} plus a very
small electron scale $\delta_e$ where the electron frozen-in
constraint is broken. We did not include $\delta_e$ in this
calculation to simplify the analysis and because $\delta_e$ does not
appear to substantially modify the reconnection rate in well developed
Hall mediated reconnection\cite{shay98b,Hesse01}.

{\bf Simulations:} Eqns.~\ref{dndt}-\ref{dbdt} with $z_h = 1$ and the
same normalizations were integrated forward in time using F3D, a
parallel fluid code. The simulation domain is a uniform grid of $2048
\times 1024$ grid points with the physical size $L_x \times L_z =
204.8 \times 102.4,$ with periodic boundaries at $x = \pm L_x/2$ and
$z = \pm L_z/2.$ The initial equilibrium consists of a system size
double current sheet with $B_x = B_0 \{\tanh[(z+L_z/4)/w_0] -
\tanh[(z-L_z/4)/w_0] - 1\}$ and $w_0 = 1.5.$ ${\bf V}_h = 0$ initially
with $n_h = 0.64$ everywhere and $T_i = T_e = T_h = 0.5.$ Pressure
balance is maintained by setting $B^2/2 + (T_i + T_e) n_i = B_0^2/2 +
(T_i + T_e) n_{i0},$ where $B_0 = 1.0 $ and $n_{i0} = 1.0$ are the
values outside the current sheet. A final equilibrium constraint is
$n_i V_{iy} = J_y\,T_i/(T_i + T_e),$ which determines $V_{iy}.$ The
remainder of the equilibrium current is put into $V_{ey}.$ In order to
break the frozen-in constraint of the electrons at the smallest
scales, the term $\mu_4 \nabla^4 {\bf B}$ has been added to the RHS of
Eq.~\ref{dbdt}, with $\mu_4 = 5 \cdot 10^{-5}$. To initialize the
double tearing mode, x-lines were seeded in both current sheets at
$(x,z)$ = $(\pm L_x/4,\mp L_z/4)$ with an initial half island width
$w=0.55.$ A small amount of random noise was added to the initial {\bf
B} and ${\bf V}_i$ of about $10^{-3}$ their equilibrium values.

To examine the effect of the heavy ion mass, We ran three simulations
with $\hat{m} = \{1,16,10^4\}$ and $n_h = 0.64$ in all cases. The
first case corresponds to the usual two-species reconnection. The
second case corresponds to reconnection in the presence of O$^+$, with
$\{d_s,d_i\sqrt{n_i n_e/(z_h n_h)^2},d_h\} = \{0.8,2,5\}.$ In the
third case, the 3 length scales are $\{0.8,2,125\},$ so that the heavy
ions form an immovable background and the global scales are controlled
by the heavy whistler.

\begin{figure}
\epsfig{file=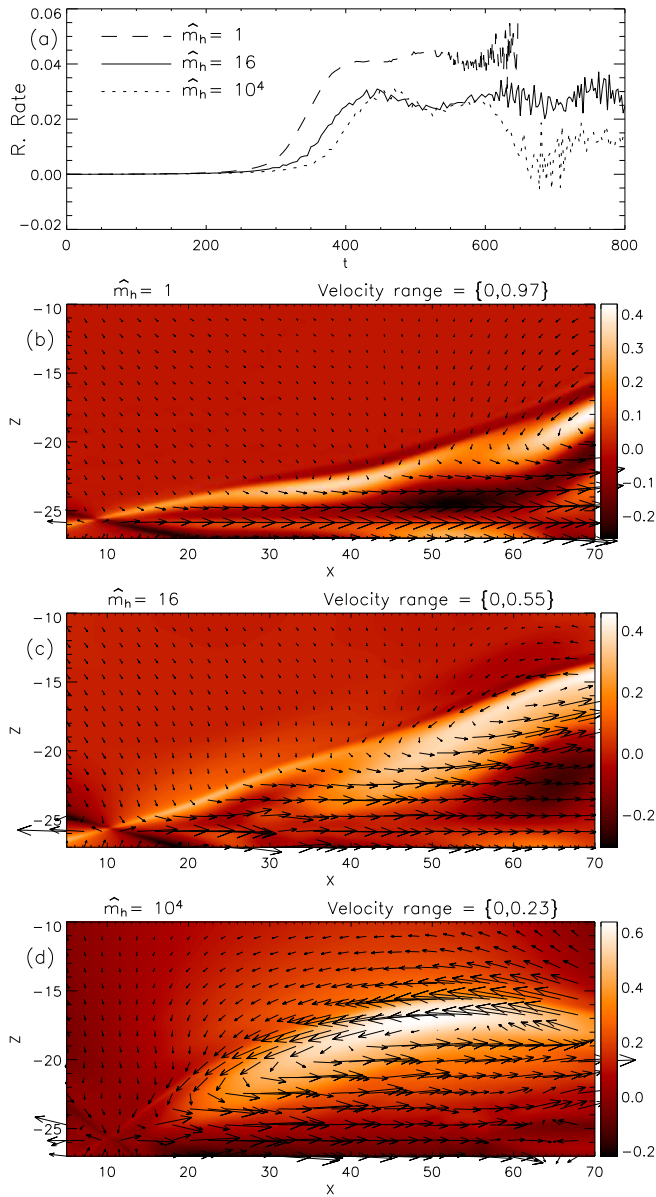,width=8.7cm}
\caption{(Color online) (a) Reconnection rates. (b)-(d) $B_y$ with
proton flows, (b) $\hat{m}_h = 1$ and $t = 500,$ (c) $\hat{m}_h = 16$
and $t=650,$ (d) $\hat{m}_h = 10^4$ and $t=650.$}
\label{overview}
\end{figure}

The reconnection rates of the three simulations versus time are
shown in Fig.~\ref{overview}a. 
The $\hat{m}_h = 1$ case clearly shows a substantially larger
reconnection rate. The two cases with heavy ions show very similar
reconnection rates, but the largest $\hat{m}_h$ shows a large decrease
in its reconnection rate around $t=600,$ while the $\hat{m}_h=16$ case
keeps a steady rate. The heavy whistler velocity has a k dependence,
$V \sim k d_h c_{Ah}.$ As reconnection proceeds in a system and the
island width $w$ gets larger and larger, the effective $k \sim 1/w$
for the reconnection process decreases. Because the heavy whistler is
mediating global convection in the $\hat{m}_h = 10^4$ case, as the
global convection scale length increases, the global convection
velocity must decrease, throttling the reconnection rate.

The reconnection generates very different signatures for the different
$\hat{m}_h.$ Figs.~\ref{overview}b-d show the out-of-plane $B_y$
generated from the reconnection and the proton flow vectors. The
x-line is located close to $(x,z) = (10,-25.6)$ in all three
cases. Only a small fraction of the total simulation is shown. The
$\hat{m}_h = 1$ case shows the usual quadrupolar structure generated
by frozen-in electron flow\cite{Mandt94}. For $x > 45,$ the clean
quadrupolar signature begins to change to a more complicated structure
with both positive and negative $B_y$ due to the finite system
size. Because $V_{ix}$ is maximum at about $x=45,$ and for greater $x$
the slow-down of $V_{ix}$ causes a compression of $B_z,$ and the
resulting $J_y$ generates a $B_y$ signature of the opposite sign. 

\begin{figure}
\epsfig{file=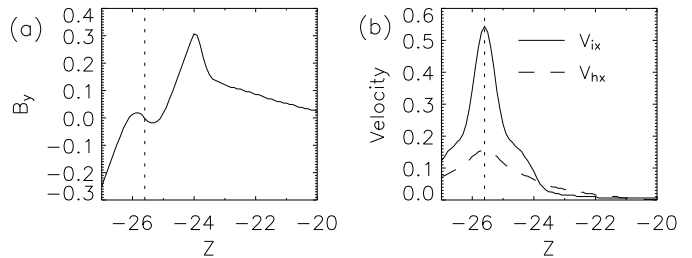,width=9cm}
\caption{For the O$^+$ case, a cut along $z$ at $x=20.0:$ (a) $B_y$ and
(b) $V_{ix}$ and $V_{hx}$. The vertical dotted line is the symmetry
axis ($z=-25.6$).}
\label{slice.light}
\end{figure}

The $\hat{m}_h = 16$ case (Fig.~\ref{overview}c) shows signatures of
both the light and heavy whistler. For $x < 40$ a narrow band of
positive $B_y$ associated with the light whistler is present. A cut of
$B_y,$ $V_{ix},$ and $V_{hx}$ at $x=20$ is shown in
Fig.~\ref{slice.light}.
This $B_y$ spike has a main length scale of about a $d_i,$ which is
roughly consistent with the light whistler cut-off scale of $d_s =
0.8$ for this simulation. Like the usual two-species whistler, the in
plane current generating this $B_y$ is due to counterstreaming
parallel electron beams upstream and downstream of the $B_y$
perturbation. There is a long tail of $B_y$ upstream of the spike ($z
> -23.5$) in Fig.~\ref{slice.light}a, though, which is not present in
the two-species case. The proton outflow shows a peak on the symmetry
axis like the two-fluid case, and its velocity is much larger than the
O$^+$ velocity.

\begin{figure}
\epsfig{file=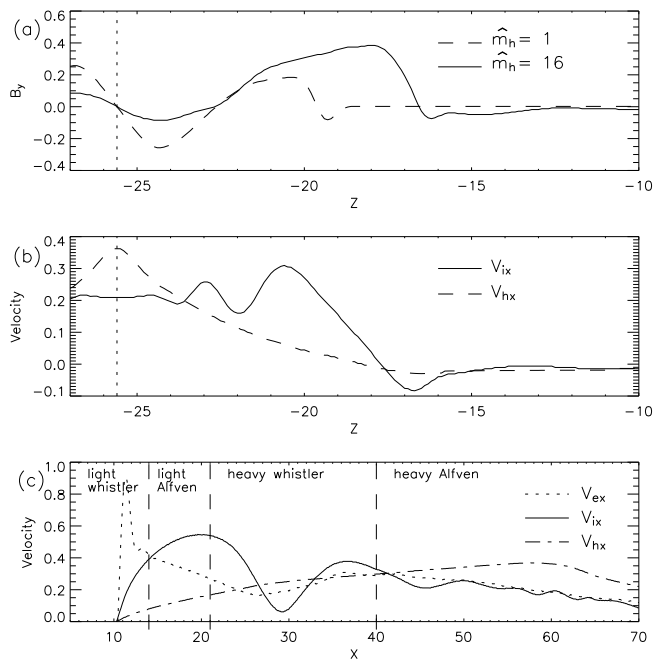,width=9cm}
\caption{(a) Slice of $B_y$ along $z$ at $x=55$, (b) slice of
$x$-velocities along $z$ at $x=55$ for $\hat{m}_h = 16,$ (c) slice of
$x$-velocities along $x$ at $z=-25.6$ for $\hat{m}_h = 16$. }   
\label{slice.heavy}
\end{figure}

The quadrupolar $B_y$ becomes dominated by the heavy whistler for $x >
40$ in the $\hat{m}_h = 16$ case. The $B_y$ signature broadens out
substantially because $d_h = 5$ for this simulation, and the current
which generates it is carried by both the ions and
electrons. Fig.~\ref{slice.heavy}a shows comparison slices for the
$\hat{m}_h = 1$ and $16$ cases at $x=55.0.$
The main positive $B_y$ spike is substantially wider in the $\hat{m}_h
= 16$ case, although it is not 5 times wider as might be expected from
a comparison of $d_h$ to $d_i.$ The $x$-velocities reveal another key
signature, as shown in Fig.~\ref{slice.heavy}b for $\hat{m}_h = 16.$
The parallel ion flows from the heavy whistler associated with $B_y$
lead to a negative $V_{ix}$ at about $z = -17.$ Also, the ion flow no
longer is maximum at the symmetry line, but instead peaks off axis at
around $z = -20.5.$ On the symmetry line, $V_{hx}$ is somewhat larger
than $V_{ix}.$ $V_{ix}$ is about 4 times slower in the $\hat{m}_h =
16$ case than in the $\hat{m}_h = 1$ case. The off axis peak of
$V_{ix}$ and the substantial negative $V_{ix}$ (about 1/3 of
maximum ion outflow) do not occur unless the heavy whistler is active.

In the case with $\hat{m}_h = 10^4,$ the heavy ions are immovable and
the heavy whistler is dominant at the global scales in the simulation
as seen in Fig.~\ref{overview}d. The main peak of $B_y$ is quite wide,
and there is a nonzero $B_y$ out to global scales. The parallel ion
flows which flow with the Hall electron currents are much stronger in
this case.

The multiscale structure of the dissipation region is demonstrated for
the O$^+$ case in a cut of the outflows away from the x-line, as shown
in Fig.~\ref{slice.heavy}c, which is a cut along $x$ through the
center of the current sheet.  For $x < 21,$ the behavior of the flows
is very similar to that seen in two-species
reconnection\cite{Shay99,Shay04}. In the light whistler region, the
electrons accelerate to speeds much faster than the ions and exceeding
the relevant Alfv\'en speed. The electrons cannot maintain this high
velocity and are forced to decelerate to a speed comparable to the
protons when they reach the light Alfv\'en region. Inside of this
Alfv\'en region, the protons reach their maximum velocity. The
protons, in an analogy to the electrons in the light whistler region,
cannot flow this speed indefinitely, and are forced to slow down
inside the heavy whistler region. Finally, their speed becomes
comparable to the O$^+$ outflow in the heavy Alfv\'en region. Inside
the heavy whistler region at $x=29,$ however, $V_{ix}$ drops nearly to
zero, below $V_{hx}.$ This behavior also occurs in the $\hat{m}_h =
10^4$ case at about $x=18,$ as seen in Fig.~\ref{overview}d. The O$^+$
outflows behave like the proton outflows in the two-fluid case,
gradually accelerating and finally reaching their maximum velocity in
the outermost Alfv\'en region. 

{\bf Discussion} As discussed in the introduction, a Sweet-Parker like
analysis of the dissipation region yields $V_{in} \sim (\delta/D)
c_{At}$. For the $\hat{m}_h = \{1,16\}$ cases, we would expect the
outflow speeds to differ by a factor of $\sqrt{(m_i n_{i16} + m_h
n_h)/(n_{i1} m_i)} = 2.6$ between the two cases, where $n_{i1} = 1.64,
n_{i16} = 1.0,$ and $n_{h} = 0.64.$ The maximum outflow in the
two-species case is about 1.0, while for the case with O$^+$ is .35,
giving a ratio of 2.9, quite close to what is expected. If $\delta/D$
stayed the same between the two simulations, the O$^+$ case would be
expected to reconnect nearly 3 times slower than the two-species
case. In Fig.~\ref{overview}a, however, the two cases asymptote to
approximately steady-state rates that differ by around 1.5, leaving a
factor of about 2 unaccounted for. A rough estimation of the scaling
of $\delta/D$ between the simulations may be possible by examining the
angle, $\theta,$ that the $B_y$ signature makes with the $z=-25.6$
symmetry line. Presumably $\delta/D \sim \tan \theta.$ This gives
$\tan \theta_i = 0.08$ and $\tan \theta_{{\rm O}^+} = .15,$ which
sheds light on the factor of two difference. A more careful
determination of $\delta/D$ as well as a scaling study with very large
system sizes will be necessary to determine if this change in
$\delta/D$ is robust.

These results imply that a substorm occurring with a high enough O$^+$
density ($m_h n_h \gg m_i n_i$) will have slower outflows and a
reduced reconnection rate normalized to the upstream proton Alfv\'en
speed. With all else being equal, this implies that the expansion
phase of substorms will take longer to occur or will reconnect less
lobe field in the same amount of time. However, substantial O$^+$
populations tend to occur during times of increased geomagnetic
activity. The magnetotail equilibrium, and thus the lobe magnetic
fields and density, may be modified substantially during these
periods, which may offset or even overpower the reduction in the
reconnection rate.

{\bf Acknowledgments} The authors thank E. Zweibel and L. Rudakov for
valuable discussions. This work was supported in part by NASA and the
NSF. Computations were carried out at the National Energy Research
Scientific Computing Center.


\end{document}